\documentstyle[preprint,aps]{revtex}
\begin{document}
\draft
\preprint{\vbox{\hbox{UT-Komaba/97-3}\hbox{hep-th/9701117}\hbox{(To appear
in Nucl.Phys.{\bf B})}}}

\baselineskip= 18 truept

\def\w{\wedge}
\def\e{\epsilon}
\def\ra{\rightarrow}
\def\a{\alpha}
\def\b{\beta}
\def\n{\eta}
\def\g{\gamma}
\def\d{\delta}
\def\t{\theta}
\def\o{\Omega}
\def\l{\lambda}
\def\p{\phi}
\def\P{\Phi}
\def\cg{\cal G}
\def\cb{\cal B}
\def\pr{\partial }
\def\bpr{\bar {\pr }}
\def\ca{{\cal A}}
\def\cl{{\cal L}}
\def\cm{{\cal M}}
\def\zb {{\bar z}}
\def\na{{\nabla }}
\def\s{{\sigma}}
\def\G{{\Gamma }}
\def\l{{\lambda }}
\newcommand{\be}{\begin{equation}} \newcommand{\ee}{\end{equation}}
\newcommand{\bea}{\begin{eqnarray}}\newcommand{\eea}
{\end{eqnarray}}

\title{\bf D-Branes and Twelve Dimensions} 

\author{\large{Supriya Kar}
\footnote{e-mail: supriya@hep1.c.u-tokyo.ac.jp }}

\address{{\vspace{.1in}}
{{Institute of Physics,}\\
{University of Tokyo, Komaba, Tokyo 153, Japan}}}


\maketitle

\thispagestyle{empty}

\begin{abstract}
\baselineskip= 14 truept

We study the D-brane solutions to type IIB superstring in ten dimensions
and find interpretation in terms of compactification of a twelve
dimensional three-brane of (a specific) F-theory on a torus $T^2$. 
In this frame-work, there also exist a two-brane which may be argued 
to be equivalent to the three-brane by utilizing the electric-magnetic
duality in eleven dimensions. In this context, we propose for the
existence of an isometry in one of the transverse directions to the
three-brane in F-theory. As a consequence the two-brane may be identified
with the three-brane in twelve dimensions itself. The twelve dimensional
picture of D-branes in type IIB theory suggests for the reformulation of
type IIB superstring in terms of three-brane of F-theory.

\end{abstract}

\vfil

\newpage

\section{Introduction}

In the recent years, string dualities \cite{sen,dkl,sch,ms,s,as,ht,witten}
can be understood by assuming the existence
of higher dimensional theories. There are several connections between 
superstring theories in various dimensions \cite{witten,dlp,dlm,kmp,dmw}
and it is conjectured that all the string theories are
different phases of a single underlying theory in eleven or twelve 
dimensions. At present, the eleven dimensional quantum theory is 
know as M-theory in literature \cite{s} whose low energy description
is believed to be the eleven dimensional supergravity theory \cite{cjs}.
On the other hand, the twelve dimensional theory known as F-theory
\cite{vafa} is understood as a class
of type IIB superstring compactifications containing $7$-branes where 
the dilaton and the Ramond-Ramond ($RR$) scalar (axion) vary on the 
internal manifold \cite{vafa,witten96,mv,ferrara,senO,bds,hull,kv}.
Thus the $8$-dimensional
world-volume of $7$-branes of type IIB string on a D-manifold $S^2$ is 
interpreted as the compactification of F-theory on $K3$ which is in turn dual 
to heterotic string theory on a two dimensional torus $T^2$. 
The duality between F-theory on $K3$ and type IIB on $T^2$ has been 
established near the orbifold limit of $K3$ \cite{senO}. 
In this context, we consider a (specific) 
F-theory in twelve dimensions \cite{vafa,witten96,mv,ferrara} 
on a torus $T^2$
which is an artifact underlying the 
$SL(2,Z)$ invariance \cite{s,as,ht,witten} of the type IIB theory in
ten dimensions. We analyze the D-brane solutions to the bosonic sector
of type IIB in ten dimensions and interpret their origin in the three-brane
solutions to the low-energy limit of a hypothetical F-theory. In order to
arrive at a plausible unified picture of three-brane in twelve dimensions 
itself, we propose for the existence of an isometry in one of the transverse 
directions to branes. 

\vspace{.1in}
Among various other recent developments of string theory, the extended  
solutions namely p-branes with RR charges have played an important 
role in understanding the non-perturbative effects \cite{fllq} required
by string duality \cite{pol}. These objects have been studied considerably
and within open string theory known as Dirichlet branes (D-branes). For a
review see ref.\cite{pcj}. It is observed that at low energies below
string scale, the dynamics of D-branes is governed by the effective
theory of massless modes of the open strings whose end points lie on the
$(p+1)$-dimensional worldvolume of the D-(p)brane. In fact the open string 
state is described by Dirichlet boundary conditions for $(9-p)$-transverse
directions and Neumann boundary conditions for $(p+1)$ worldvolume directions.
In a simple case of a single $p$-brane, the effective theory is the reduction
of a ten dimensional $N=1$ supersymmetric abelian gauge theory to 
$(p+1)$ dimensional world-volume of the brane \cite{wit}.
The interaction between D-branes by means of open string calculations 
have been studied in the current literatures \cite{wit,gm,dkps}.
Interestingly enough, the scattering of closed string states from a
quantized D-particle has been discussed recently \cite{hk}.

\vspace{.1in}
In type II theories, RR charges are associated with $(p+1)$ form
potentials which are carried by the D-(p)branes of the theory. These include
the p-branes for $p=0,2,4,6,8$ in type IIA and $p=-1,1,3,5,7,9$ in type IIB 
theories. In this context, various D-brane and their $T$-duals have been 
analyzed \cite{bho}. The exact type IIB RR backgrounds have also been addressed
using geometrical arguments \cite{kks}. It is known that, only D-(p)branes with
$0\leq p\leq 6$ have a natural Minkowski space interpretation as the dual
of a p-brane in $D=10$ dimensions is a $(6-p)$ brane. Nonetheless,
type IIB $7$-brane has a $(-1)$-brane 
dual which are sources of the RR charge and has an interpretation as 
$D$-instanton \cite{ggp}. Furthermore the dual of type IIA $8$-brane
\cite{brgpt} is interpreted as a cosmological constant in massive type
IIA supergravity and type IIB $9$-brane can be viewed as $D=10$ space-time.
Generally, the non-triviality in these
$D(p)$-brane solutions is due to the harmonic function of the transverse 
$(9-p)$ spatial coordinates defining the radial direction which is due to the
$BPS$ nature of the solution. It is observed that for $p\leq 6$ the  
harmonic functions are asymptotically flat and for $p\geq 7$ the
asymptotic properties are of different nature. The $p$-volume tension of
D-(p)brane is found to be ${1\over{g_s}}$ as they are $BPS$ saturated 
\cite{witten}. Thus in the weak coupling limit (${g_{s}}<< 1$) 
$D$-branes are localized, however extremely small
compared to Nevue-Schwarz Nevue-Schwarz (NS-NS) $p$-branes and their dynamics
have been studied \cite{dkps}. As a consequence
the non-perturbative objects ($D$-branes) are considered to be intermediate
between the fundamental string and solitonic 
five brane. In addition to D-branes, 
an understanding of M-branes have been analyzed \cite{pt} and a
super-matrix formulation of M-theory has been suggested \cite{town}.
In this context a matrix model underlying the 
M-theory \cite{bfss} and type IIB theory \cite{ikkt} have been proposed. 

\vspace{.1in}
In the recent past there are various interesting proposals underlying the 
twelve dimensional origin of type IIB theory \cite{gr}. In an interesting paper
\cite{tseytlin}, the D-instanton of type IIB theory is viewed as a wave in 
twelve dimensions. It is known that the field equations of the
type IIB theory \cite{schwarz} can not be
obtained from a ten dimensional covariant action. However with vanishing
five-form (self-dual) field strength the equations of motion can be derived
from an action. The bosonic sector of the type IIB is known to
possess an exact global $SL(2,Z)$ invariance \cite{s,as,ht}. Under this ten
dimensional $SL(2,Z)$ $S-$duality, the complex scalar (say $\l $) formed out
of an RR scalar ($\chi $) and the 
dilaton ($\P $); $i.e.\l = \chi + i e^{-\P}$ undergoes a fractional linear
transformation and is identified as the modulus of a torus $T^2$. 
The two two-form
potentials, one in the NS-NS sector and the other in the RR sector get
exchanged and the metric along with
the four-form potential remain invariant. In order
to understand the $SL(2,Z)$ invariance of type IIB in ten dimensions, one
compactifies M-theory on a torus to nine dimensions and type IIB
on a circle. Then $SL(2,Z)$ of type II theory in nine dimensions gets
interpreted as the symmetry of the torus \cite{s,as}. 
However, only in the limit of zero
torus area one obtains a type IIB in ten dimensions. On the other hand if
one starts with twelve dimensions and compactifies on $T^2$, then
$SL(2,Z)$ gets interpreted as the symmetry of the torus in ten
dimensions \cite{vafa}. It has been 
conjectured that a D-string (defined with a
world-sheet gauge field) has a critical dimension twelve with metric
signature ($2,10$) and is 
related to ten dimensional theory by null reduction \cite{ov}. 

\vspace{.1in}
In this paper, we consider a hypothetical 
dynamical theory in twelve dimensions by lifting
a three-form gauge field and a four-form one from an eleven dimensional 
M-theory and a ten dimensional type IIB theory respectively. Our analysis  
is in the spirit of F-theory as an underlying theory for type IIB superstring.
We study the D-branes in type IIB theory and find their gravitational
counterpart in a specific F-theory. In this frame-work, the three-brane and 
the six-brane (electric-magnetic dual of two-brane) in F-theory can be
identified with the M-(two)brane and its dual M-(five)brane respectively in
eleven dimensions. However, the three-brane does not seem to be 
directly identified with the two-brane in F-theory.
Thus, with a motivation for an unified picture of three-brane
in F-theory, we need to propose for the existence of a spatial isometry in
one of the transverse directions. As a consequence the six-brane may be
redefined and a D-(five)brane in type IIB  theory can be viewed as the
wrappings of a seven-brane in F-theory  on a torus. This is consistent
with the observation 
that with torus compactification of the twelve dimensional theory, the
modulus of the type IIB string theory in ten dimensions can be identified with
the modular parameter of the torus. We find that there exist $p$-branes 
($p=1,2,\dots ,7$) with singularities and
they can be shown to be related to three-brane by duality
symmetries in twelve dimensions itself.
Thus, our analysis suggests that D-branes (D-strings, self-dual three-brane
and D-(five)brane) in type IIB string theory may be
viewed as the wrappings of a three-brane on a torus in F-theory.

\vspace{.1in}
We organize the paper as follows. In section 2, we discuss about the
construction of a $D=12$ classical theory in the spirit of ref.\cite{ferrara}
and write down the dynamics of the background fields \cite{tseytlin}. We
show that the $D=12$ theory when compactified on a torus $T^2$ gives rise
to a ten dimensional theory similar to type IIB string. 
We begin section 3, with a D-string of type
IIB theory and explicitly show that it can be viewed as the wrappings of 
three-brane on a torus $T^2$ in F-theory. We find charged two-brane 
solution to the low-energy limit of F-theory and identify that with the 
three-brane solution by proposing an isometry in one of the transverse
directions. In section 4, we present a self-dual three-brane to type IIB string
theory and view that as the wrappings of five-brane on a torus in F-theory.
In section 5, we study a D-(five)brane of type IIB string theory and 
identify with the six-brane in F-theory by proposing an isometry in 
one of the transverse directions.  
Finally in section 6, we summarize our results and conclude with 
a note that the D-branes in type IIB superstring theory may be viewed as
the wrappings of three-brane on a torus in F-theory.

\section{Twelve dimensions}
	
It is know that the $SL(2,Z)$ invariance in the field equations of type IIB
in $D=10$ provides hints about the existence of twelve dimensional theory.
However there is no supergravity theory in $D=12$. Inspite of the fact, it
may be possible to discuss a hypothetical twelve dimensional theory
(may be with some kind of fermionic symmetries) with additional 
constraints on the background fields. Recently supersymmetry in $D=12$ has
also been discussed \cite{hp}. With the present understanding, it is
believed that the $D=12$ theory should contain type IIB fields in $D=10$
along with those of M-theory in $D=11$. It is observed that the four-form
potential (say $D_4$) of type IIB supergravity (zero-slope limit of type IIB
superstring) and the four-form field strength (say ${\tilde F}_4$) of $D=11$
supergravity (low-energy limit of M-theory) are lifted to $D=12$ dimensions
in order to accommodate both type IIB three-brane and M-theory five-brane. As
a result, it is natural to view three-brane and five-brane electric-magnetic
duality in $D=12$. Thus, one finds an electric M-(five)brane as a magnetic
three-brane in F-theory and vice-versa.

\vspace{.1in}
In order to account for the possible fields and their dynamics 
in twelve dimensions, we start with
the geometrical coupling \cite{ferrara} and write down 
the Chern-Simons term :

\be
S_{CS} = \int_{{\cal M}_{12}} {\hat D}_4 \wedge {\hat F}_4 \wedge {\hat F}_4
= \int_{{\cal M}_{12}} {\hat C}_3 \wedge {\hat F}_4 \wedge {\hat F}_5\;\ ;
\ee
where ${\hat D}_4$ is the four-form potential 
(${\hat F}_5=d{\hat D}_4$) and ${\hat F}_4=d{\hat C}_3$ is
the four-form field strength defined with a three-form potential ${\hat C}_3$
in $D=12$. Henceforth we use $hat$ on the fields to denote them in
$D=12$. 

\vspace{.1in}
In fact there can be one 
more Chern-Simons coupling which is topological in nature
by construction and is given by
\be
T_{CS}= \int_{{\cal M}_{12}} {\hat F}_4 \wedge {\hat F}_4 \wedge {\hat F}_4
\;\ .
\ee
With an assumption ${{\cal M}_{11}}=\pr {\cal M}_{12}$, the topological
coupling (2) can be identified with that of M-theory. However the geometrical
coupling (1) may be identified with that of M-theory and type IIB theory 
after compactifications on a circle $S^1$ and a torus $T^2$ respectively
without such assumption. It is noted that the geometrical coupling in eq.(1)
induces currents ${j^*}_3={\hat F}_4\wedge {\hat F}_5$ 
and ${j^*}_4=(1/2){\hat F}_4\wedge {\hat F}_4$ 
for the gauge fields ${\hat C}_3$ and
${\hat D}_4$ respectively. In presence of five and six brane sources (say
${\hat q}_5$ and ${\hat q}_6$), these currents are not conserved. However,
redefining the currents appropriately it is shown \cite{ferrara} 
to be conserved. 

\vspace{.1in}
To begin with the dynamics for the $D=12$ theory \cite{tseytlin}, we 
streamline our discussions with the motivation for the F-theory as an
underlying theory for type IIB string theory \cite{vafa}. We also utilize the 
fact that two of the two-form potentials in type IIB theory form a
doublet under an $SL(2,Z)$ transformation. It is known that there  
can (only) be one-form potentials on a torus $T^2$ which originate
from a three-form in $D=12$ and there is only one four-form potential
in $D=10$. Thus, it is natural to expect two two-form
potentials forming an $SL(2,Z)$ doublet in $D=10$ from the dimensional 
reduction of the $D=12$ three-form potential
${\hat C}_3$ on $T^2$. The four-form potential ${\hat D}_4$ is assumed to
be that of $D=10$ and with necessary constraints assumed to describe a 
self-dual 
three-brane. As a consequence, one arrives
at the Chern-Simons coupling in type IIB theory from that of $D=12$ in eq.(1) 
on $T^2$. In principle there may be
other higher form of potentials in $D=12$. However
for our purpose, 
we consider the truncated theory describing three-branes and 
five-branes. Also, the action should contain a scalar curvature
$R^{(12)}$ and in this case we define it with a metric of signature $(2,10)$
\cite{vafa}.
Following the discussions, it is plausible to write down the action
describing the hypothetical theory in $D=12$ \cite{tseytlin} as

\be
S_{12D} \equiv {\int }_{{\cal M}_{12}}d^{12}x\ {\sqrt {-\hat G}}\Big ( 
R^{(12)} - {1\over{2.4!}} {{\hat F}_{4}}^2 - 
{1\over{2.5!}} {{\hat F}_{5}}\Big )
+{\int}_{{\cal M}_{12}}
{\hat D}_4 \wedge {\hat F}_4 \wedge {\hat F}_4 + \dots \;\; .
\ee
Let us consider a compactification on $T^2$ with signature $(1,1)$,  
so that the $D=10$ metric has the Minkowski signature $(1,9)$. We restrict
the background configuration in $D=12$ by defining the fields
in terms of that in $D=10$ as :

\vspace{.1in}
\noindent the metric $\hat G$ with signature ($2,10$) \cite{vafa}

\be
{\hat G} = \pmatrix { {G} & {0} \cr {0} & {M} \cr }\;\ ;
\ee
where G represents the metric with ($1,9$) signature in $D=10$
and M with ($1,1$) is the $2\times 2$ matrix
parameterizing the moduli fields in an $SL(2,Z)$ invariant form on a
torus $T^2$. In our notation, the expression for the moduli field 
$M$ in terms of scalars (the axion $\chi$ and the dilaton $\P$ ) 
is given by

\be
M= \pmatrix {{e^{\P}} & {\chi e^{\P}} \cr {\chi e^{\p}} &
{\chi ^2 e^{\P} + e^{-\P}} \cr }\;\ ;
\ee
where the modulus $\l=\chi +ie^{-\P}$ and det $M=1$.
We write the three-form potential

\be
\hat C = \pmatrix {{0} & {B^q} \cr {B^p} & {0} \cr }
\ee
and the four-form potential as

\be
\hat D = \pmatrix {{D} & {0} \cr {0} & {0} \cr }\;\ .
\ee
Where $B^p, B^q$ for $p=1,2$ and $D$ 
are the two-form potentials and a four-form one in ten dimensions respectively.

\vspace{.1in}
Now, we write the effective action in ten dimensions by compactifying the one
in twelve dimensions (3) on a torus $T^2$ as :

\bea
S_{10D}\equiv {\int }_{{\cal M}_{10}} d^{10}x {\sqrt {-G}}&& \Big [ R^{(10)}+
{1\over4} Tr \Big ( {\pr}_{\a}M{\pr}^{\a}M^{-1}\Big )\nonumber\\ 
&&-{1\over{2.3!}}
H^{(p)}M_{pq}H^{(q)} 
-{1\over{2.5!}}{F_5}^2 \Big ] 
+ {\int}_{{\cal M}_{10}} D_4 \wedge dB^{1}\wedge dB^{2}\;\ ;
\eea
where the metric $G$ is in Einstein frame defined with ($1,9$) signature and
$H^{p}=dB^{p}$ for $p=1,2$ are the $SL(2,Z)$ invariant field strengths 
corresponding to the two-form potentials. The scalar curvature $R^{(10)}$
and the moduli field $M$ consisting of two scalars ( identified with 
the axion $\chi$ and the dilaton $\P$ in eq.(5)) are also $SL(2,Z)$ 
invariant forms and obtained directly from
the curvature $R^{(12)}$ using the compactifications in eq.(4). The five-form
field strength in eq.(8) is not a self-dual one as required by the type IIB
superstring theory. However, modulo the five-form field strength $F_5$  the
action in eq.(8) may be identified with that of type IIB theory without the
self-dual field strength in $D=10$.
Since, there is no covariant form of
the type IIB action in $D=10$, we identify the equations
of motion corresponding to eq.(8) with the type IIB equations \cite{schwarz}
by assuming necessary constraints on the background fields in $D=12$.
In principle the self duality of the five-form
field strength $F_5$ and the supersymmetry in $D=10$ have to come
naturally from $D=12$.
These issues need deeper understanding of the subject and 
remain  unanswered in the present frame-work 
of $D=12$. However the analysis of classical solutions in $D=12$
representing the gravitational counter part of the well established 
D-branes of type IIB string is nevertheless interesting. We analyze
the wrappings of charged branes on a torus in F-theory and hope that
our results may shed some light in understanding the conjectured F-theory.

\vspace{.1in}
In a different context, it is shown \cite{ferrara} that the
three and four-form potentials are not independent in twelve dimensions 
in order to obtain a type IIB self-dual field strength. It is also
plausible to consider (only) three-brane and its electric-magnetic dual
five-brane in the frame-work of twelve dimensional theory of gravity due
to natural reason. However in this frame-work, we find two-brane solutions 
corresponding to a three-form potential in addition to the
three-brane corresponding to a four-form one in F-theory. In
fact, we find that a three-brane in F-theory can be identified with a 
M-(two)brane in $D=11$ which is dual to a M-(five)brane. Furthermore, the 
M-(five)brane can be lifted to $D=12$ and may be viewed as a six-brane which
is dual to the two-brane in F-theory. Thus following a via route through
$D=11$, it is possible to argue for an unified picture describing a three-brane
in F-theory. On the other hand in the frame-work of $D=12$ itself, we show that
the two-brane of F-theory can be identified 
with the three-brane by proposing an
isometry in one of the transverse directions.

\vspace{.1in}
In order to find connections between the classical solutions to eq.(3) in 
$D=12$ and that of eq.(8) in $D=10$, we write the invariant distance
in $D=12$ as :

\be
d{\hat s}^2({\cal M}_{12}) = ds^2({\cal M}_{10}) + ds^2(T^2)
\ee
where ${\cal M}_{12}$ is the Minkowski space of signature ($2,10$) \cite{vafa}
in $D=12$, ${\cal M}_{10}$
is the one with the standard signature $(1,9)$ 
in $D=10$ and the $2nd$ term represents 
the line element on $T^2$
with the metric $(1,1)$ defined in eq.(5). In our discussions of classical
solutions, we consider the metric in $D=12$ with ($2,10$) signature. However
a physical signature ($1,11$) can be obtained by Wick rotating one of the
time coordinates.

\vspace{.1in}
In a recent work \cite{tseytlin}, it is shown that a D-instanton of type IIB
appears as a gravitational wave in $D=12$. A D-instanton in type IIB theory
can be described by the RR scalar (axion $\chi$) which has its origin in
the scalar curvature $R^{(12)}$ in $D=12$. We present 
the classical solution in $D=12$ corresponding to a D instanton in our
notations as :

\be
d{\hat s}^2 = -\Big (1-{{{Q}_{-1}}\over{{\bf r}^8}}\Big ) dt^2 
+ \Big (1 + {{{Q}_{-1}}\over{{\bf r}^8}}\Big ) dy^2
- {{2{{Q}_{-1}}}\over{{\bf r}^8}} dydt
+ d{\bf r}^2 + {\bf r}^2 d{\o_9}^2\;\ ;
\ee
where $t$ is the time-like coordinate and $y$ is the spatial coordinate
defined on a torus. The radial coordinate
$\bf r$ is defined over the ten of the transverse coordinates. The global
electric charge ${Q}_{-1}$ is the D-instanton charge and  can be interpreted
as a linear momentum in $D=12$. The signature of the metric is ($1,11$) which
may be obtained by rotating one of time coordinates ($t\rightarrow iy$) 
in a metric with ($2,10$) signature. Upon compactification on a torus with
signature ($1,1$), the $D=10$ describes an euclidean space ($0,10$).

\vspace{.1in}
Thus a D-instanton of type IIB theory has a gravitational counterpart in 
$D=12$. The case of $7$-branes of type IIB theory has already been addressed in 
ref.\cite{vafa}. In the subsequent sections, we analyze and discuss about 
the $D=12$ gravitational counterparts of 
D-string, its electric-magnetic dual D-(five)brane 
and the self-dual three-brane of type IIB string theory.

\section{D-string as a three-brane}

In this section, we begin with a D-string solution \cite{gm,bho}
to the type IIB equations motion in $D=10$ \cite{schwarz}. We set
the $RR$ scalar $\chi $, 
the self-dual five-form field strength $F_5$ and 
the three-form field strength
$H$ in the NS-NS sector as well as all the
fermionic fields to zero. Then we write down the the extremal 
D-string configuration consisting of the metric ($1,9$), 
the two-form potential $B_2$ in the RR sector and the dilaton $\P$ as :

\bea
ds^2 = &&{{H_1}({\bf r})}^{-{3\over4}} 
\Big [ -dt^2 + dy^2\Big ] + H_1({\bf r}) 
\Big [ d{\bf r}^2 + {\bf r}^2 d{{\o }_7}^2\Big ]\;\; ,\nonumber\\
&& B_2 = \pm {{H_1}({\bf r})}^{-1}\;\ , \nonumber\\ 
&& e^{\P} = {{H_1}({\bf r})}^{1\over2}\nonumber\\
{\rm with}\qquad\qquad && H_1({\bf r})=1+ {{q_1}\over{{\bf r}^6}}\;\ ;
\eea
where $y$-coordinate is parallel to string and 
defines the worldsheet along with the time-like coordinate $t$. The radial
coordinate ${\bf r}$ : 
${\bf r}^2 ={\delta}_{ij}x^ix^j$ is defined with the orthogonal
coordinates ($i,j= 1,2,\dots ,8$) to string and the angular part 
described by $d{\o_7}^2$ is the $SO(7)$ invariant line element on $S^7$. 
The global electric charge $q_1$
corresponds to the two-form potential $B_2$ in the
$RR$ sector. $H_1(\bf r )$ is the harmonic function of transverse 
coordinates and the metric  has a singularity at ${\bf r}=0$.

\vspace{.1in}
Now consider a $D=12$ theory in eq.(3) and  we are interested to obtain
a three-brane solution. It is natural to expect that the D-string in 
type IIB may appear as the wrappings of twelve dimensional three-brane 
on a torus $T^2$. We restrict the background configurations by setting
the four-form field strength to zero (${\hat F}_4=0$)
and solve for the equations of motion in eq.(3). The consistent background 
configuration consists of the metric, the four-form potential ${\hat D}_4$
and can be given as :

\bea
{d{\hat s}^2} = &&{{{\hat H}_3}({\bf r})}^{-{1\over2}} 
\Big [-dt_adt^a + dy_b dy^b\Big ] +
{{\hat H}_3({\bf r})}^{1\over3} 
\Big [d{\bf r}^2 + {\bf r}^2 d {{\o }_7}^2\Big ]
\nonumber\\
&&{\hat D}_4 = \pm {{\hat H}_3({\bf r})}^{-1}
\nonumber\\
{\rm with} \qquad\qquad &&{\hat H}_3({\bf r}) 
= 1+ {{{\hat q}_3}\over{{\bf r}^6}}\;\ ;
\eea
where $t_a\; ; a=1,2$ are the time-like coordinates and $y_b\; ;b=3,4$ 
are the spatial coordinates defining the worldvolume of signature 
($2,2$) for the three-brane. The radial coordinate
${\bf r}$ is defined on the transverse
plane over rest of the eight coordinates. The last term in eq.(12)
describes the $SO(7)$ invariant line element on $S^7$. The conserved
electric charge ${\hat q}_3$  
corresponds to the four-form potential
${\hat D}_4$ in $D=12$. The harmonic function ${\hat H}_3(\bf r )$ is 
defined with the transverse coordinates and can be shown to possess 
a point singularity at ${\bf r}=0$.
The solutions in eq.(12) describes an electrically charged three-brane
in $D=12$. By Wick rotating one of time-like coordinates and then 
compactifying on one of the spatial coordinates on the world-volume, the
three-brane in F-theory may be identified with M-(two)brane in $D=11$.
This is a consequence of the fact that the M-(two)brane corresponding
to the four-form gauge field is lifted to $D=12$. In the asymptotic
limit ($i.e. {\bf r}\rightarrow \infty $) the three-brane becomes flat.

\vspace{.1in}
\noindent
Rescaling the D-string metric in eq.(11) with respect to the dilaton $\P$ 

\be
 G'=e^{-{{4\P}\over3}} G \;\; ,
\nonumber\\
\ee 
we rewrite the metric as :
\be
{ds^2} = {H_1}{(\bf r)}^{-{2\over3}}
{{H_1}({\bf r})}^{-{3\over4}}\Big [-dt^2+ dy^2\Big ] 
+ {{H_1}({\bf r})}^{1\over3}\Big [ d{\bf r}^2 + {\bf r}^2 d {\o _7}^2\Big ]
\;\ .
\ee
Since $SL(2,Z)$ invariance of the field equations of type IIB string
gives rise to a theory in $D=12$, it should contain a $D=10$ metric
and the other two dimensions arise  from a torus with the
modular parameter $\l$ defining the moduli matrix $M$. Following the
discussions in the previous section, we lift the D-string configurations
to $D=12$ by writing down the metric as :

\bea 
d{\hat s}^2= {H_1}{(\bf r)}^{-{2\over3}}
{{H_1}{(\bf r)}}^{-{3\over4}}\Big [-dt^2 && + dy^2\Big ]
- {{H_1}{(\bf r)}}^{1\over2} d{z_1}^2  
\nonumber\\
&& + {{H_1}{(\bf r)}}^{-{1\over2}}d{z_2}^2 +
{{H_1}{(\bf r)}}^{1\over3}\Big [ dr^2 + r^2 d{\o_7}^2\Big ]
\eea
where $z_1$ is the time-like and $z_2$ is the spatial coordinate on 
the torus. Redefining the worldsheet coordinates $(t,y)$ along with $z_1$   
appropriately, we identify the background metric (15) with that of
$D=12$ in eq.(12). In the process the two-form potential $B_2$ in eq.(11)
is lifted to
$D=12$ and is identified with the four-form potential ${\hat D}_4$ in eq.(12).
Note that the harmonic function ${H_1}({\bf r})$ in eq.(11) is identical to
that of ${\hat H}_3(\bf r )$ in eq.(12). The D-string electric charge $q_1$
gets interpreted as a three-brane charge ${\hat q}_3$ in $D=12$. Thus we
obtain an electrically charged three-brane (12) with  world-volume signature
($2,2$) in $D=12$ starting from a D-string (11) in $D=10$. In other words, a
three-brane in $D=12$ after double dimensional reduction on 
$T^2$ gives rise to a D-string of type IIB. Our analysis suggests 
that the D-string in type IIB theory may have its origin in F-theory 
and can be viewed as the wrappings of a charged three-brane on a torus.
In addition, the three-brane in F-theory may also be viewed as a M-(two)brane
in $D=11$. Then following an electric-magnetic duality in $D=11$, M-(two)brane
can be identified with the M-(five)brane which may be lifted to $D=12$ to 
represent a six-brane and finally identified with the two-brane in F-theory.
However in $D=12$ itself the two-brane does not seem to be related to 
three-brane in F-theory.

\vspace{.1in}
To understand the nature of two-brane in $D=12$, we
consider the case with vanishing four-form potential ${\hat D}_4=0$ and
with non-zero three form potential 
${\hat C}_3$. We solve for the equations of motion (3) and obtain the
background configuration consisting of the metric and the three-form potential
${\hat C}_3$ as :

\bea
{d{\hat s}^2} = &&{{\hat H}_2({\bf r})}^{-{2\over3}}
\Big [-dt^adt_a + dy^2 \Big ] +
{{\hat H}_2({\bf r})}^{2\over7} \Big [d{\bf r}^2 + 
{\bf r}^2 d {{\o }_8}^2\Big ]\;\; ,
\nonumber\\
&& {\hat C}_3 = \pm {{H_2}({\bf r})}^{-1} 
\nonumber\\
{\rm with}\qquad\qquad &&{\hat H}_2({\bf r}) 
= 1+ {{{\hat q}_2}\over{{\bf r}^7}}
\eea
where $t_a\;\ ;a=1,2$ are the time-like coordinates and $y$ is the
spatial coordinate defining the world-volume for the 
two-brane. The radial coordinate {\bf r} is defined on the transverse
plane with nine orthogonal coordinates and $d{\o_8}^2$ corresponds to the
$SO(8)$ invariant line element on $S^8$. The conserved electric
charge ${\hat q}_2$ corresponds to the three-form potential
${\hat C}_3$ in $D=12$. The solution in eq.(16) describes an electrically
charged two-brane in $D=12$ with a point singularity at ${\bf r}=0$ and is
asymptotically flat.

\vspace{.1in}
In order to have an unified picture in $D=12$, it is
essential to relate the two-brane (16) with the three-brane in F-theory.
As a consequence, D-string in type IIB theory can also be viewed as a 
two-brane in F-theory. With the motivation, 
we propose for the existence of an isometry in one of the 
transverse directions to that of two-brane in $D=12$. Then the harmonic 
function ${\hat H}_2({\bf r})$  can be reduced to a function of eight of the
transverse coordinates instead of nine. The modified harmonic function
can be written as
$$ {{\hat H}'}_2({\bf r})=1+ {{{\hat q}_2}\over{{\bf r}^6}}\;\ . $$
Then the metric in eq.(16) can be rewritten as :

\be
{d{\hat s}^2} = {{{{\hat H}'}_2}
({\bf r})}^{-{2\over3}}\Big [-dt_adt^a + dy^2 \Big ] 
+ dx^2 + {{{{\hat H}'}_2}({\bf r})}^{2\over7}
\Big [ d{\bf r}^2 + {\bf r}^2 d {\o _7}^2\Big ]\;\ .
\ee
The three-form potential retains its expression in eq.(16) with the
modified harmonic function as defined above. The metric in eq.(17) along
with the gauge field in eq.(16) describes 
a charged two-brane and can be shown to be identified with the three-brane
in F-theory. The two-brane may also be related to M-(two)brane by a 
spatial compactification in $x$-direction.

\vspace{.1in}
\noindent
In order to view the D-string in eq.(11) directly as a two-brane
in F-theory, we rescale the D string metric as : 

\be
G''=e^{-{{10\P}\over7}} G\;\ .
\ee
Following similar arguments as before, we lift the D-string solution in
eq.(11) with the rescaled metric in eq.(18). Redefining the worldsheet 
coordinates appropriately, we identify the D-string backgrounds in $D=12$ 
with that of two-brane (17) in F-theory. 
In this case, the electric charge of D-string can also be identified with
that of two-brane charge in $D=12$. Similarly, it is possible to identify
the two-brane backgrounds directly with the three-brane in F-theory once
an isometry is established in one of the orthogonal directions to the 
two-brane.

\vspace{.1in}
We summarize our results in this section with the  observation that a
D-string in type IIB theory can be viewed as the wrappings of three-brane
on torus in F-theory. By proposing an isometry in one of the transverse
directions, the two-brane in F-theory is shown to be identified with the
three-brane in $D=12$. As a result, the three and four-form gauge fields
describing the two and three-brane respectively in F-theory are not independent
of each other. This is consistent with the observation \cite{ferrara}
where it is argued to obtain a self-dual field strength in
type IIB theory from $D=12$. Our proposal for isometry leading to the 
identification of a two-brane with the three-brane in $D=12$ itself suggests 
for an unified picture of three-brane in F-theory. It may be shown that 
a four-brane arises from a three-brane in F-theory due to the proposed 
isometry in one of the transverse directions to the three-brane world-volume.

\section{Self-dual three-brane as a five-brane}

In this section, we show that the self-dual three-brane in IIB string theory
appears as a five-brane in $D=12$ which is infact reduces to a type
IIB theory on a torus $T^2$. Thus for the purpose, we consider a type IIB
superstring and restrict to self-dual three-brane with vanishing one-form
potentials along with the RR scalar and the fermionic sector. 
The background configuration consists of the metric, the self-dual four-form
potential $D_4$ and the dilaton $\P$. Solving the type IIB equations of motion
\cite{schwarz}, one obtains the extremal solution \cite{bho} as :

\bea
ds^2 = && {H_3({\bf r})}^{-{1\over2}} \Big [ -dt^2 + d{y_s}dy^s\Big ] + 
{H_3({\bf r})}^{1\over2} 
\Big [ d{\bf r}^2 + {\bf r}^2 d{{\o }_5}^2\Big ]\;\; ,
\nonumber\\
&& D_4 = \pm {H_3({\bf r})}^{-1}\;\ ,
\nonumber\\
&&\P= const. \nonumber\\
{\rm with}\qquad\qquad 
&&H_3({\bf r})=1+ {{q_3}\over{{\bf r}^4}}
\eea
where $t$ is time-like and $y_s\; ;s=1,2,3 $ are the spatial coordinates
defining a three-brane world-volume of signature ($1,3$). The radial 
coordinate ${\bf r}$ is defined with the transeverse
coordinates $x^i$ ($i,j= 1,2,\dots ,6$) to D-(three)brane and the last 
term represents the $SO(5)$ invariant line element on $S^5$. The 
charge $q_3$ corresponds to the  self-dual four-form 
potential $D_4={D^*}_4$ and thus carry both electric and magnetic charge
simultaneously. The extremal backgrounds in eq.(19) describing a self-dual
three-brane may be identified to that in ref.\cite{hs}.

\vspace{.1in} 
Let us consider the twelve dimensional theory (3) with non-zero five-form field 
strength and vanishing four-form field strength ${\hat F}_4=0$. We dualize
the field strength and obtain the required seven-form field strength 
${{\hat F}^*}_7$
to describe a five-brane in $D=12$. 
We solve for the background fields containing the metric and six-form
potential ${{{\hat D}^*}_6}$ (3) and obtain :

\bea
d{\hat s}^2 =&& {{\hat H}_5({\bf r})}^{-{1\over3}}
\Big [ -dt_adt^a + d{y_b}dy^b\Big ] + 
{{\hat H}_5({\bf r})}^{{1\over2}} 
\Big [ d{\bf r}^2 + {\bf r}^2 d{{\o }_5}^2\Big ]\;\ ,
\nonumber\\
&& {{\hat D}^*}_6 = 
\pm {{\hat H}_5({\bf r})}^{-1} \nonumber\\
{\rm and} \qquad\qquad
&&{\hat H}_5({\bf r})=1+ {{{\hat q}_5}\over{{\bf r}^4}}
\eea
where $t_a\; ;a=1,2$ are the time-like and $y_b\; ;b=3,4,5,6$ are the spatial
coordinates parallel to the five-brane defining the world-volume. The
radial coordinate ${\bf r}$ is
defined of the transverse coordinates 
($i,j= 1,2,\dots ,6$) orthogonal to the five-brane and the angular part is
described by ${\o_5}$ on $S^5$. The topological magnetic charge ${\hat q}_5$
corresponds to the six-form potential ${D^*}_6$ in (20). The solution
in eq.(20) describes a magnetically charged five-brane in $D=12$ and the
metric can be shown to possess a point singularity at ${\bf r}=0$. The 
five-brane solution in $D=12$ is also asymptotically flat.

\vspace{.1in}
In order to identify the self-dual three-brane with 
the five-brane in $D=12$ on $T^2$, we lift the $D=10$ solution in eq.(19) to
$D=12$ by adding two extra dimensions on a torus following (9)  
which is in fact expressed 
in terms of moduli matrix (5) in $D=10$. However in the case of self-dual
three-brane, the moduli fields are constants. Thus, it is straightforward
to lift the metric in eq.(19) to $D=12$ and can be given by

\be
d{\hat s}^2={H_3(\bf r )}^{-{1\over3}}\Big [ -dt^2 + dy_s dy^s\Big ]
-d{z_1}^2 + d{z_2}^2
+ {H_3(\bf r)}^{1\over2}\Big [ d{\bf r}^2 + {\bf r}^2d{\o_5}^2\Big ]\;\ ;
\ee
where $z_1$ is the time-like and $z_2$ is the spatial coordinates
defined on the torus. 
Redefining the the world-volume coordinates along with $z_1$ and $z_2$
appropriately, we arrive at the background metric in $D=12$ as :

\be
d{\hat s}^2 =
{H_3({\bf r})}^{-{1\over3}}\Big [ -dt_adt^a + d{y_s}d{y^s}
\Big ] 
+{H_3({\bf r})}^{1\over2} \Big [ d{\bf r}^2 + {\bf r}^2 d{\o_5}^2 \Big ]
\;\ .
\ee
Comparing the harmonic function in eq.(19) with that of eq.(20),
we notice that self-dual charge $q_3$ may be identified  with the
magnetic charge ${\hat q}_5$. Then the three-brane metric in eq.(22)
along with the four-form potential and dilaton in eq.(19) may be  
identified with the five-brane solution in eq.(20).
This analysis suggests that a self-dual three-brane may
be viewed as a five-brane in a $D=12$ theory and the self-dual 
three-brane charge plays the role of a five-brane magnetic 
charge. In other words, a self-dual three-brane with world-volume
signature ($1,3$) may appear as the wrappings of five-brane
($2,4$) signature on a torus in $D=12$. Using the electric-magnetic duality
the self-dual three-brane may also be related to the electric three-brane
in F-theory.

\section{D-(five)brane as a six-brane}

In this section, we analyze the  D-(five)brane to type IIB string theory
and show that it  can be viewed  as a six-brane in twelve dimensions by
proposing an isometry in one of the transverse dimensions. We consider the 
D-(five)brane solution \cite{bho} to type IIB string equations of motion 
\cite{schwarz} by setting the two-form potential to zero in the NS NS 
sector ($B=0$) along with that of four-form potential
($D_4=0$) and the scalar ($\chi=0$) in RR sector. Thus, the only
non-vanishing background fields are the metric, the six-form
potential ${B^*}_6$ (which is dual to the two-form one) and the dilaton
$\P$. The extremal 
D-(five)brane configurations describing the above backgrounds can
be obtained by solving the type IIB equations of motion and can be given as :

\bea
ds^2 = && {H_5({\bf r})}^{-{1\over4}}\Big [ -dt^2 + d{y_s}dy^s\Big ] + 
{H_5({\bf r})}^{3\over4} 
\Big [ d{\bf r}^2 + {\bf r}^2 d{{\o }_3}^2\Big ]\;\; ,
\nonumber\\
&& {B^*}_6 = \pm {H({\bf r})}^{-1}\;\ ,
\nonumber\\
&& e^{\P} = {H_5({\bf r})}^{-{1\over2}}\nonumber\\
{\rm with} \qquad\qquad &&H_5({\bf r})=1+ {{q_5}\over{{\bf r}^2}}
\eea
where $t$ is the time-like and $y_s\; ; s= 1,2,\dots ,5$ are the 
coordinates parallel to 
the five-brane defining the six-dimensional world-volume.
The radial coordinate ${\bf r}$ is
defined with the orthogonal coordinates $x^i$
($i,j= 1,2,\dots ,4$) to D-(five)brane and $d{\o_3}^2$ represents the
invariant line element on $S^3$. The topological magnetic charge $q_5$
corresponds to the six-form 
potential ${B^*}_6$ in the RR sector.

\vspace{.1in}  
Now consider the twelve dimensional theory (3) with vanishing five-form field
strength ${\hat F}_5=0$ and non-zero four-form field strength. In this section,
we are interested to find six-brane which is  dual to two-brane in $D=12$.
Six-brane is described by a seven-form potential with an eight-form
field strength (${{\hat F}^*}_8$). We solve for the background fields (3)
consisting of metric and the seven-form potential ${{\hat C}^*}_7$ 
and obtain

\bea
d{\hat s}^2 = && {{\hat H}_6({\bf r})}^{-{2\over7}}
\Big [ -dt_adt^a + d{y_b}dy^b\Big ] + 
{{\hat H}_6({\bf r})}^{2\over3} 
\Big [ d{\bf r}^2 + {\bf r}^2 d{{\o }_4}^2\Big ]\;\; ,
\nonumber\\
&&{{\hat C}^*}_7 = \pm {{\hat H}_6({\bf r})}^{-1} 
\nonumber\\
{\rm with}\qquad\qquad &&{\hat H}_6({\bf r})=1+ {{{\hat q}_6}\over{{\bf r}^3}}
\eea
where $t_a\; ;a=1,2$ are the time-like and $y_b\; ;b=1,2,\dots ,5$
are the spatial coordinates parallel to 
the six-brane defining the world-volume with ($2,5$) signature. As before
the radial coordinate ${\bf r}$ :
${\bf r}^2 ={\delta}_{ij}x^ix^j$ 
is defined on the rest of the transverse coordinates
$x^i$ ($i,j= 1,2,\dots ,5$) to six-brane. ${\hat q}_6$
is the topological magnetic charge corresponding to the seven-form 
potential ${{\hat C}^*}_7$. It can be shown that the magnetically charged
six-brane metric has a point singularity at ${\bf r}=0$ and is asymptotically
flat. The six-brane solution in eq.(24) when compactified on one of the spatial
coordinates of world-volume may be identified with that of M-(five)brane
in $D=11$. Further a M-(five)brane is dual to a M-(two)brane and following 
discussions in the section 3, the six-brane can be identified with a 
three-brane in F-theory. In fact this is consistent with the observation
\cite{ferrara} where it is argued that the five-brane of M-theory is lifted 
and may be identified with the three-brane of F-theory. 

\vspace{.1in}
In order to view the D-(five)brane in eq.(23) as a six-brane in eq.(24)
in $D=12$ itself, we consider a spatial isometry in one of the transverse 
directions to the six-brane world-volume. As a result the harmonic function
is modified and dependent on four of the transverse coordinates instead of
five and  can be given by 
$$
{{{\hat H}'}_6({\bf r})} = 1 + {{{\hat q}_6}\over{{\bf r}^2}}. 
$$
The metric in eq.(24) can be rewritten separating the symmetry axis
(say $x$) from the rest of the transverse directions and is given by

\be
d{\hat s}^2 = {{{\hat H}'}_6({\bf r})}^{-{2\over{7}}} 
\Big [-dt^2 + d{y_b}dy^b\Big ] + dx^2 +
{{{\hat H}'}_6({\bf r})}^{2\over3} 
\Big [ d{\bf r}^2 + {\bf r}^2 d{\o_3}^2 \Big ] \;\ .
\ee
Redefining the world-volume coordinates along with the spatial coordinate
$x$, the six-brane may be viewed as a seven-brane in $D=12$. In order to
identify the D-(five)brane in eq.(23) with the six-brane backgrounds consisting
of the potential (24) and the metric (25), we  rescale the  D-brane metric as :

\be
G'= e^{{\P}\over6}G \;\ .
\nonumber\\
\ee
Following eq.(9) we lift the D-brane metric and rewrite in $D=12$ as :

\bea
d{\hat s}^2 = {{H}_5({\bf r})}^{-{1\over{3}}} 
\Big [-dt_adt^a && + d{y_s}dy^s\Big ] 
- {{H_5}(\bf r)}^{-{1\over2}} d{z_1}^2\nonumber\\
&& 
+ {{H_5}(\bf r)}^{1\over2} d{z_2}^2 +
{{\hat H}_5({\bf r})}^{2\over3} 
\Big [ d{\bf r}^2 + {\bf r}^2 d{\o_3}^2 \Big ] 
\eea
where $z_1$ is time-like and $z_2$ is the spatial coordinate on the torus.
Redefining the world-volume coordinates along with $z_1$ and $z_2$, we 
identify the metric in eq.(27) along with the expression for gauge field 
in eq.(23) with that of eqs.(25) and (24) respectively. Thus a  
D-(five)brane with world-volume signature ($1,5$) 
appears as a six-brane ($2,5$) in $D=12$ with an isometry in one of the
transverse directions. The topological magnetic charge of D-(five)brane
can be identified with that of six-brane. 

\vspace{.1in}
We summarize this section with the observation that a D-(five)brane
may be viewed as a six-brane in $D=12$ by proposing an isometry in one
of the transverse directions to the six-brane world-volume. In other words,
a D-(five)brane appears as the wrappings of seven-brane on a torus in F-theory.
There also exist one-brane solutions which are electric-magnetic dual to
seven-brane in $D=12$. The analysis suggests that D-(five)brane may 
be viewed as a three-brane in F-theory using the proposal for isometry along
with the duality symmetries.

\section{Discussions}

In this paper, we have considered a twelve dimensional theory
of gravity with four-form and five-form field strengths as 
an underlying theory for the type IIB superstring. The compactification of
the twelve dimensional theory on a torus is presented and the modular
parameter of the torus is identified with the modulus describing a 
complex scalar field in ten dimensions. It is argued that the resulting
equations of motion in ten dimensions may be identified with those of
type IIB superstring by constraining the background fields appropriately
in twelve dimensions. 

\vspace{.1in}
By analyzing a D-string solution to type IIB theory in $D=10$
along with a three-brane solution in $D=12$, it is shown that
D-string appears as a three-brane in F-theory and the corresponding
electric charges are identified. Thus the wrappings of 
three-brane in $D=12$ around the internal torus gives rise to a D-string of
type IIB theory. It is argued that the three-brane in F-theory can be
identified with a M-(two)brane in $D=11$ by compactifying one of the
coordinates on the world-volume in this frame-work. Two-brane solutions are 
also obtained in $D=12$ by solving the metric equation with the three-form 
gauge field. It is noticed that the two-brane and three-brane in F-theory 
do not seem to be identified with each other. However proposing the existence 
of an isometry in one of the transverse directions to the two-brane 
world-volume, it is shown to be identified with a three-brane in F-theory. 
Thus the proposal for an isometry allows one to have an unified picture
of three-brane in $D=12$ in this frame-work. Our analysis is also consistent
with the observation \cite{ferrara} that the three and four-form gauge fields
in $D=12$ are not independent of each other in order to have a self-dual
four-form gauge field in $D=10$.

\vspace{.1in}
In this frame-work, it is also shown that 
a self-dual three-brane can be viewed as a five-brane in $D=12$. 
In the process of lifting a $D=10$ theory to $D=12$, the self-dual charge
gets interpreted as a magnetic five-brane charge. Using the electric-magnetic
duality, it is argued that a self-dual three-brane can also be viewed as 
an electrically charged three-brane in F-theory. We note that in the process
of identifying the lifted self-dual three-brane with the five-brane, we do not
need to rescale the three-brane metric (the dilaton is a constant) unlike
the D-string and D-(five)brane cases. The identification of background fields
in this case may allow one to think that the five-brane in F-theory is a direct
consequence of the self-dual three-brane in Einstein frame.

\vspace{.1in}
Finally, we have shown that a D-(five)brane when lifted to $D=12$ can be
viewed as a six-brane by proposing the existence of an isometry in one of the
transverse directions to six-brane world-volume. Then the
magnetic charge of D-(five)brane can be interpreted as that of six-brane. 
At first sight this case seems to be in a different footing than
the other two cases where the D branes are viewed as the wrappings
of higher branes on an internal torus in $D=12$. However with the existence
of an isometry, a six-brane can be related to a seven-brane in $D=12$ and
a similar picture for the D-(five)brane may be viewed. A six brane is dual
to a two-brane in $D=12$ and in turn may be identified with a three-brane
in F-theory. By compactifying one of the time-coordinates
of the six-brane world-volume, it can be shown to represent a
M-(five)brane in $D=11$. This is a consequence of the fact that the $D=12$
theory of gravity is constructed by taking into account the M-(five)branes
in $D=11$ along with the self-dual three-brane of type IIB theory. Thus
the analysis suggests that a magnetically charged M-(five)brane appears as
an electrically charged three-brane and a self-dual three-brane 
as a five-brane in F-theory which are indeed electric-magnetic duals in $D=12$.

\vspace{.1in}
We have not succeeded in listing the constraints on
the background metric and gauge fields in $D=12$, so that one obtains 
the desired 
type IIB fields in $D=10$ directly upon dimensional reduction on $T^2$. Among 
various other issues, the origin of supersymmetry in type IIB remain
unanswered in this frame-work of twelve dimensional theory of gravity. Since
the extremal solutions to the type IIB theory are viewed as the three-brane in 
the low-energy limit of F-theory, it may be interesting to find out the
corresponding fermionic symmetries in $D=12$.
We note that with the existence of an isometry there are
additional charged $p$-branes in $D=12$. Nonetheless, it is possible to relate
all the $p$-branes ($p=1,2\dots ,7$) in $D=12$ itself due to the existence of
an isometry. Our analysis suggests that the 
D-branes of type IIB string theory
may be reformulated and in turn can be viewed as a three-brane in F-theory.

\vspace{.35in}

\noindent{\bf Acknowledgements :}

\vspace{.15in} 
I would like to thank Professor Yoichi Kazama for various comments and 
useful discussions. I am grateful to the high energy physics theory group
in the Institute for various help and support. Also, I wish to thank M.J.
Duff and S. Hewson for useful correspondences. This work is supported
by the Japan Society for the Promotion of Science, JSPS-P96012.

\def\np{{\it Nucl. Phys.} {\bf B}}
\def\pl{{\it Phys. Lett.} {\bf B}}
\def\pr{{\it Physics Reports }}
\def\prl{{\it Phys. Rev. Lett.}}
\def\prd{{\it Phys. Rev.} {\bf D}}
\def\ijmp{{\it Int. J. Mod. Phys.} {\bf A}}
\def\mp{{\it Mod. Phys. Lett.} {\bf A}}

\end{document}